# From Coulomb Blockade to Resonant Transmission in a MoS$_2$ Nanoribbon

*Yanjing Li, Nadya Mason\**

University of Illinois at Urbana-Champaign, Frederick Seitz Materials Research Laboratory

ABSTRACT: We have measured a nanoribbon of MoS$_2$ at low temperature, and observed the evolution of the system from a regime of multiple small quantum dots in series to one where the entire nanoribbon acts as a single quantum dot. At higher Fermi energies, resonant transmission through disorder-induced potential wells is evident. Our findings shed light on the length scale of quasi-ballistic transport in the material.

Although the transition metal dichalcogenide molybdenum disulfide (MoS$_2$) has been studied for decades[1,2], recent advances in nanoscale materials characterization and fabrication have substantially extended the range of possibilities in creating nanostructures and devices from it. In particular, it has been shown that MoS$_2$ can be exfoliated into two-dimensional layers of single unit cell thickness, and that both monolayer and bilayer MoS$_2$ have a significant energy gap[3-6]. Devices based on thin layers of MoS$_2$ have demonstrated strong photoluminescence[7,8], a current on/off ratio exceeding 10$^8$ in field-effect transistors (FETs)[9], and efficient valley and spin control by optical helicity[10-12]. By making materials not only thinner but also narrower, significant confinement effects as well as other quantum properties are expected to appear. For example, samples with confinement on a mesoscopic scale may exhibit Coulomb blockade or Fabry-Perot oscillations, which depend on the contact transparency and mean free path, and which provide information about coherent transport that could be useful in quantum device applications[13]. In addition to ballistic transport properties, disorder-induced localization plays an important role in materials having confinement on a mesoscopic scale[14, 15]; thus, understanding the effect of disorder on transport is crucial to scaling down and utilizing nanodevices. An advantage of studying nanoribbons is accessing this interplay between phase coherence, environmental disorder and mean free path. Beyond this, rich physics and applications have been predicted specifically for $MoS_2$ nanoribbons, such as band gap modification, ferromagnetism, and metal-insulator transition tuning with a transverse electric field[16]. However, while the electronic properties of thin layers of $MoS_2$[17] have been studied via transport, the properties of $MoS_2$ with geometry confinement have not been well-studied[18], particularly at low temperatures where quantum effects are relevant. Here, for the first time, we demonstrate the transport properties of a $MoS_2$ nanoribbon device at low temperature. We observe a gate-tunable transition from Coulomb blockade to resonant transmission, where the transition point occurs when the entire nanoribbon acts as a quantum dot. Our observations show that mesoscopic confinement effects can dominate transport in these small structures. The results also reveal the length and energy scales at which quasi-ballistic versus disorder-scattering behavior determines the transport.

Graphene nanoribbons were fabricated using commercially available crystals of molybdenite (SPI Supplies Brand Moly Disulfide), from which thin layers of $MoS_2$ were mechanically exfoliated[19] onto a Si substrate with 300 nm $SiO_2$, which serves as the back gate capacitor. The device described in this manuscript has a thickness of 1.1 nm, which corresponds to a bilayer[9]. The sample is patterned into a nanoribbon (500 nm length and 200 nm width, as can be seen in Fig. 1, lower bottom inset) and a sidegate, using electron beam lithography and reactive ion etching (RIE) with oxygen. Although there are multiple methods for making nanostructures in MoS$_2$[20, 21], we find that oxygen-based RIE does not etch the substrate and thus avoids the problem of gate leakage[18]. Source and drain leads (35 nm Ti/10 nm Au) are fabricated via another electron beam lithography and evaporation step. A micrograph of the final device is shown in Fig. 1, upper left inset.

The dc conductance of the nanoribbon, as a function of backgate potential at room temperature, shows characteristic behavior of an *n*-doped semiconductor, as shown in Fig. 1. If we assume Ohmic contacts (which allows us to estimate a lower-bound for mobility), we find that the nanoribbon has a field effect mobility ~ 0.127 cm$^2$/Vs, using $\mu = [dI_{ds}/dV_{backgate}] \times [L/(WC_gV_{ds})]$, where $I_{ds}$ is the drain current, $V_{backgate}$ is the gate voltage, $C_g$ is capacitance per unit area of 300 nm thick SiO$_2$ (12 nF/cm$^2$), $V_{ds}$ is the drain voltage, and *L* and *W* are the length and width of sample, respectively. This is consistent with typical mobility in MoS$_2$ of 0.1 ~ 10 cm$^2$/Vs [9, 22, 23], although high temperature annealing and keeping the sample in high vacuum can yield much higher mobility (60 ~ 500 cm$^2$/Vs)[24, 25]. The mobility for the nanoribbon is thus reasonable—given that it is unencapsulated, not annealed and exposed to ambient—and can be considered a lower bound (given the likely contribution from contact resistance).

Figure 2a shows the differential conductance versus backgate at 1.7 K: the conductance is strongly suppressed compared to room temperature, and has a threshold voltage at a much more positive value. These observations are consistent with the insulating behavior seen in these materials at low temperature[17, 26]. A large conductance gap is evident for 0 V < $V_{backgate}$ < 40 V and persists to large bias (50 mV), as can be seen in Fig. 2a. (The gap is also evident in the negative backgate regime, up to – 60 V). However, at backgate voltages above ~ 40 V, the conductance varies strongly with gate and bias: the 2D map of Fig. 2b shows a gap that rapidly decreases with increasing bias and backgate voltages. Just above 40 V the gap edges begin to exhibit Coulomb diamond-like features, as can be seen more clearly in the zoomed-in 2D conductance map of Fig. 3a.

The data in Fig. 2 might initially seem to indicate that the band gap suppresses conductance for gate voltages below ~ 40 V. However, a closer analysis of the data demonstrates that the band gap alone may not be sufficient to account for this large gap, and that Coulomb blockade likely plays a role as well. We find a backgate efficiency α = $\Delta V_{bias}/\Delta V_{backgate}$ = 0.02 from the resonant transmission line features marked in yellow in Fig. 2b (discussed in more detail later in the manuscript). This allows us estimate that a gap persisting up to $\Delta V_{backgate}$ = 100 V is equivalent to an energy (bias) gap of α × $\Delta V_{backgate}$ = 2 V. However, for bilayer $MoS_2$ the band gap is only 1.6 eV [4], i.e., smaller than the observed gap. In nanoribbons, the band gap could be even smaller, depending on the edge configuration[27, 28]. This suggests that the gap in this region can be dominated by other effects, in this case likely Coulomb blockade due to multiple weakly-coupled quantum dots in series.

Coulomb blockade is evident in the 2D conductance as irregular diamond patterns that vary with bias and backgate (e.g., Fig. 3a), where the average size of the diamonds corresponds to an average charging energy. At $V_{backgate}$ ~ 40 V, the charging energy (gap in bias) is ~ 50 meV. From this charging energy, the quantum dot size can be estimated as ~ 18 nm, using $E_{charging} = e^2/(8\varepsilon_0\varepsilon_r r)$, where the relative permittivity $\varepsilon_r = ((\varepsilon_{Air} + \varepsilon_{SiO_2})/2)$ and $r$ is the radius of quantum dot. This gives an upper-bound on the size of the small dots that suppress conductance for $V_{backgate}$ < 40 V. In Fig. 2b, it can be seen that as the backgate voltage increases from 40 V to 50 V, the size of the bias gap decreases, indicating that the quantum dot size increases; for example, for $V_{backgate}$ ~ 45 V, Fig. 3a shows that the average charging energy is ~ 25 meV, corresponding to a dot size of ~ 37 nm. Near $V_{backgate}$ ~ 50 V, the dot size increases to ~ 69 nm.

The Coulomb diamonds near $V_{backgate}$ ~ 50 V can be further examined by tuning the sidegate voltage, as can be seen in Fig. 4c. In this case, the charging energy is ~ 13.3 meV, which is consistent with the backgate data. As $V_{backgate}$ increases above 50 V, the charging energy decreases—implying that the size of the quantum dot increases—until it saturates at ~ 5 meV for the backgate range 55 V ~ 66 V, as shown in Fig. 2b and marked by the white dotted line in Fig. 3b. In this regime ($V_{backgate}$ > 50 V), the quantum dot size approaches the oxide thickness (300 nm), so the capacitance geometry changes from an isolated disk approximation to a parallel plate approximation, and the charging energy must be calculated by interpolating between $E_{charging}= e^2/(8\varepsilon_0\varepsilon_r r)$ and $E_{charging}= e^2/(C_g \times A)$. This allows us to estimate the quantum dot size in the saturated regime as between 0.106 μm$^2$ and 0.267 μm$^2$, which is comparable to the ribbon size (~ 0.1 μm$^2$), suggesting that at this point the entire ribbon acts a quantum dot. The fact that the charging energy is relatively constant over a large gate range in this regime is also consistent with the dot size being fixed at the device length. This implies that it is possible to have resonant quantum transport across the entire length of the nanoribbon.

The likely cause of the Coulomb blockade is charge impurities[15, 17], which modulate the local conduction band gap position ($E_C$) to create quantum dots. Figure 4a shows a schematic of how tuning the Fermi energy with respect to the underlying impurity potential affects the size of the quantum dots. At low $V_{backgate}$, the Fermi level is sitting deep inside the disorder potential and multiple small quantum dots form. When the Fermi level moves up and out of the potential dips, the quantum dots grow and can extend to the entire nanoribbon area. While resonant tunneling at localized sites have been observed in the larger $MoS_2$ samples (~a few μm$^2$)[17], here we see a complete blockade likely due to the geometric confinement of the nanoribbon.

As the backgate voltage is increased above 65 V, crossing line features resembling Fabry-Perot oscillations appear in the 2D conductance map (yellow dashed lines in Fig. 2b and Fig. 3c).

However, it is not likely that these are due to Fabry-Perot type transport: the conductance is too small ($G_{max} \sim 0.02e^2/h$) and the expected mean free path of ~ 20 nm[29, 30] too short for ballistic transport across the entire nanoribbon. In addition, it can be calculated from the slope of the lines that the resonances do not arise from states inside the conduction band[31]: transport inside the conduction band for a 2D system with a quadratic energy dispersion relationship predicts a slope of $\alpha_{band} = \Delta V_{bias}/\Delta V_{backgate} = 2C_g/(e^2 \times D)$ = 4.6e-4, where $D$ is the density of states given by $D = gm^*/2\pi\hbar^2$ (we use effective mass $m^* = 0.39m_0$[4] and degeneracy g = 4). The calculated slope is thus 44 times smaller than the slope extracted from the data of α = 0.02. While the actual capacitance could be larger due to charge traps, the density of states of these traps would have to be 100 times larger than what is usually found for the $MoS_2$-$SiO_2$ interface[32] to match the data. Similarly, the geometric confinement in the nanoribbon is not likely to change the density of states by a factor of 44.

We argue that the crossing patterns arise from resonant transmission through the disordered potential profiles; for this higher Fermi-level range, the carriers are no longer confined by the quantum dot defined by the nanoribbon size, but are still susceptible to effects of smaller disorder potentials. The disorder-induced single particle energy levels can be inferred from the crossing patterns, and are consistent with the charging energy of the Coulomb blockade in the backgate range 44 V ~ 47 V. The single particle energy levels can be calculated by measuring the distance between two adjacent resonance lines in the backgate range 66 V ~ 74 V (examples are shown in Fig. 3c, for sets of yellow and red dotted lines). Typical energy spacings are ~ 0.3 meV, 0.6 meV and 0.9 meV. Using $\Delta E \sim \pi\hbar^2/(m^* r^2)$, the energy level spacings correspond to confinement in quantum dots of size 45 nm, 32 nm and 26 nm. These quantum dot sizes are comparable to what we found earlier at $V_{backgate}$ ~ 45 V of ~ 37 nm. We note that the resonant transmission seems to occur through length scales comparable to the predicted mean free path in $MoS_2$ of ~ 20 nm[29, 30], which is consistent with coherent, ballistic transport occurring at the scale of the mean free path. Resonant transmission through a quantum well can be understood by considering that an incoming particle with energy above the potential well still undergoes scattering from the well; at certain particle energies, resonant transmission occurs and backscattering will be minimized. Alignment of the energy levels with the source/drain leads potential gives rise to crossing positively/negatively sloped resonance lines. Both Coulomb blockade and resonant transmission are observed in the intermediate regime, as shown in the backgate range 56 V ~ 61 V in Fig. 3b. Here there are still some regions showing zero conductance at low bias, which come from Coulomb blockade in quantum dot of the ribbon size, while the crossing line features which are the signature of resonant transmission show up at higher bias.

The origin of the disorder may be trapped charges at the $MoS_2$-substrate interface, as suggested for larger $MoS_2$ samples[17]. In graphene, trapped molecules at the graphene-substrate interface have been found to be the key factor contributing to charge inhomgeneity[33]. The fact that annealed and vacuum-sealed $MoS_2$ shows a band like transport (mobility relatively constant with temperature) at low temperature suggests that adsorbents from the ambient could induce significant scattering and should be considered in the picture of disorder as well[34]. Similarly, rough edges may also play a role[14, 35]: results on graphene nanoribbons—such as gaps observed in suspended devices[36] and in those on hexagonal boron nitride[37]—suggest that rough edges can contribute to conductance gaps observed in nanostructures in two-dimensional materials[38, 39].

In conclusion, we have fabricated a $MoS_2$ nanoribbon and measured its transport properties at low temperature. We observed a gate-tuned transition from Coulomb blockade to resonant transmission, determining the formation of quantum dots that range in size from < 20 nm to the extent of the entire 500 nm-long nanoribbon as gate voltage is increased. The transition can be understood in terms of an interplay between the gate-tuned Fermi energy and quantum dots created by an underlying disordered potential: as the Fermi level moves up, the barriers that isolated the quantum dots become weaker and allow resonant conduction. Mesoscopic effects and resonant transmission are evident in $MoS_2$ having a length-scale up to 200 nm, although ballistic transport and well-defined quantum dot behavior—potentially useful for quantum information devices—should be appear in even smaller nanoribbons (< 50 nm width, < 100 nm length). Our findings shed light on the length scales of quasi-ballistic transport and disorder in $MoS_2$ and can help explain related physics in these two-dimensional semiconductors, improving device performances and enabling further applications.

AUTHOR INFORMATION

**Corresponding Author**

*Email: (N. Mason) nadya@illinois.edu.

**Author Contributions**

The manuscript was written through contributions of all authors. All authors have given approval to the final version of the manuscript.

**Funding Sources**

This work was supported by the NSF-DMR-0906521

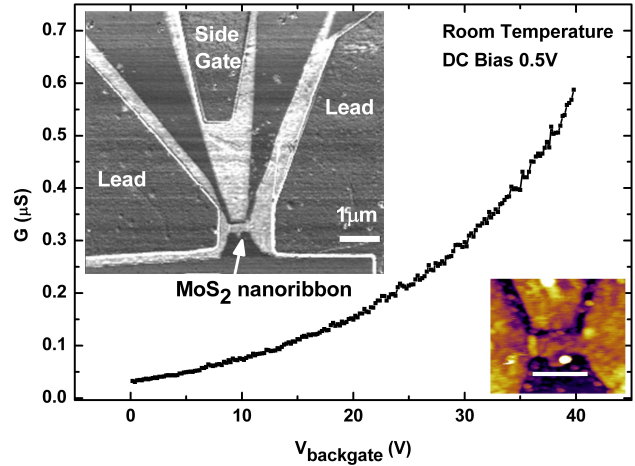

**Figure 1.** Images and room temperature characterization of the MoS$_2$ nanoribbon: DC conductance versus backgate voltage taken with a dc bias of 0.5 V. Upper left inset: AFM phase image of the device, showing the MoS$_2$ nanoribbon, source and drain leads, and the side gate. Lower bottom inset: AFM height image of the nanoribbon. Scale bar is 500 nm.

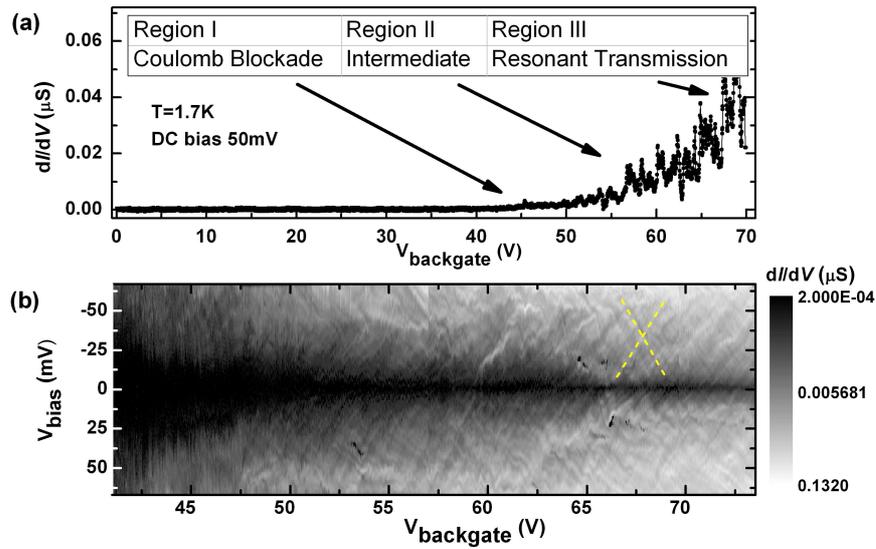

**Figure 2.** Low-temperature transport regimes: (a) Differential conductance versus backgate voltage at 1.7K with a dc bias of 50 mV and an ac excitation of 0.5 mV. (b) Two-dimensional map of differential conductance versus dc bias voltage $V_{bias}$ and backgate voltage $V_{backgate}$, with the side gate floated. Yellow dashed lines are guides to the eye showing the crossed resonance features, used to estimate the backgate efficiency. Note that (a) corresponds a line cut at $V_{bias} = $ 50mV in (b), for a larger backgate range.

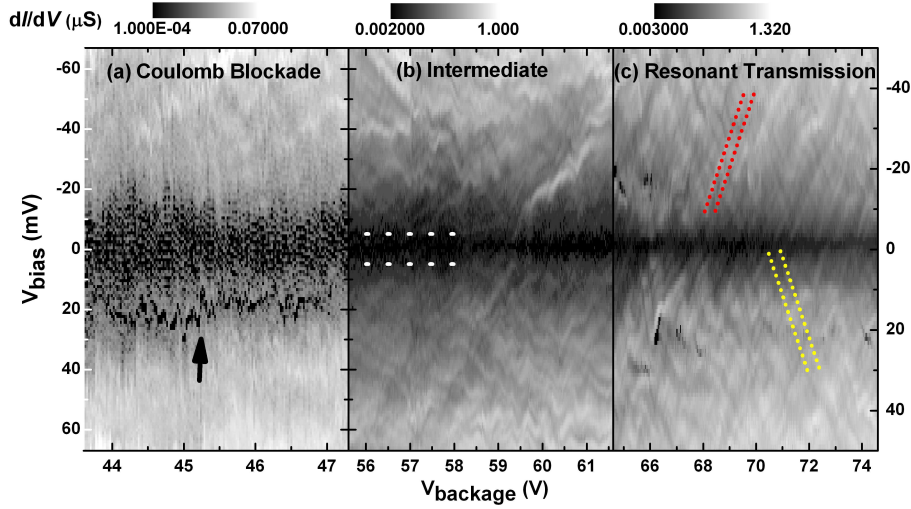

**Figure 3.** Zoom-in of regions transitioning from Coulomb blockade to resonant transmission. Each panel shows a two-dimensional map of differential conductance versus dc bias voltage $V_{bias}$ for different ranges of backgate voltage $V_{backgate}$. (a) Irregular Coulomb blockade diamonds, with black arrow marking edge of a diamond. (b) Coulomb blockade region having approximately constant charging energy ~ 5 meV. White dotted lines mark where $V_{bias}$ (± 5 meV) is equivalent to the charging energy of a quantum dot size comparable to the entire ribbon. (c) Resonant transmission region of crossed conductance patterns; typical resonances are marked by yellow and red dotted lines. The spacings between the dotted lines are used to estimate the typical energy spacing between resonant transmission energy levels.

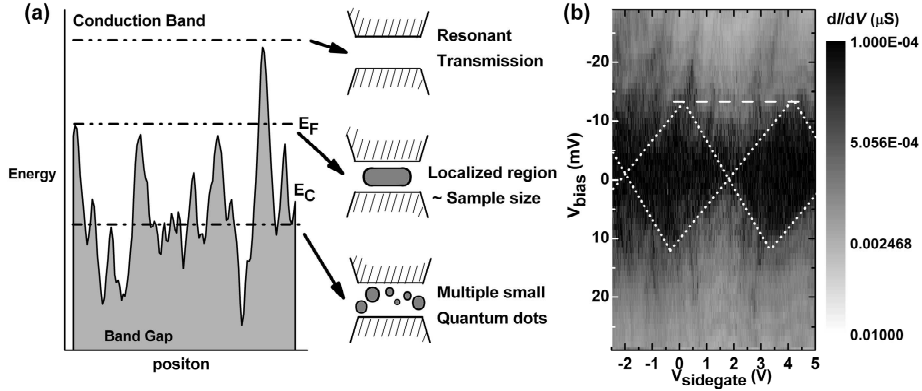

**Figure 4.** Schematic of transition from multiple dots to resonant transmission, and data demonstrating Coulomb blockade from a quantum dot, tuned by the sidegate. (a) Schematic showing how the quantum dots change in size as the Fermi level is tuned by the backgate voltage, in the presence of a disorder potential. Solid line denoted by $E_C$ shows a jagged conduction band edge modified by local disorder. The dot-dash lines denoted by $E_F$ indicate the Fermi level. When the Fermi level is deep inside the potential wells, the nanoribbon is broken into multiple small quantum dots; when the Fermi level is just above the potential wells, a quantum dot comparable to the ribbon in size emerges; when the Fermi level is further lifted up,

resonant transmission can occur due to the scattering from the potential profiles. (b) Coulomb blockade evident in two-dimensional map of differential conductance versus $V_{bias}$ and sidegate voltage $V_{sidegate}$ with the backgate voltage at $V_{backgate} = 50$V. The dotted lines outline the diamonds and the dashed line marks the charging energy at 13.3 meV.